\documentclass[a4paper, 12pt]{article}
\usepackage[dvips]{graphics}

\begin{document}

\title{\huge Scalar Particle Spectrum in broken fermion family symmetries}

\author{ F.  N.  Ndili \\
Physics Department \\
University of Houston, Houston, TX.77204, USA.}

\date{ July 2009}

\maketitle

\begin{abstract}
We study the spectrum and possible observability of Higgs-like
scalar particles associated with spontaneous breaking at high
energies of fermion family symmetries such as $SU(3)_f$ proposed
sometime ago by King and Ross. We treat the energy scale at which
the symmetry is broken as a variable not necessarily as high as
the GUT scale used by King and Ross. We compare and contrast a
non-supersymmetric treatment of the fermion family symmetry model
introduced here, and the supersymmetric treatment used by King and
Ross.
\end{abstract}

{\it{Keywords: Fermion family symmetry and the electroweak mass problem.\/}
E-mail: frank.ndili@gmail.com

\newpage

\section{Introduction}
In a number of studies, King, Ross and collaborators [1-6]  argued
that if one attributes the unsolved standard model problem of
fermion family replication, mass hierarchy, and flavor mixing, to
the existence of a continuous  family symmetry such as $SU(3)_{f}$
among the fermions, one can reproduce the observed fermion mass
patterns and flavor mixing. The family symmetry is postulated to
be exact at some high energy scale well above the electroweak
scale, but becomes  completely broken spontaneously,  before one
reaches the electroweak scale. The spontaneous family symmetry
breaking is taken to \\ occur in stages, in a manner that allows
the creation of three distinct fermion families observed  at electroweak scale. \\

For this sequential breaking of the family symmetry, King and Ross
introduced a number of new scalar fields that trigger the symmetry
breaking. This family symmetry breaking is expected to yield as
signature, a number of Higgs-like particles which  may be
observable depending on  the scale at which the symmetry is
broken. The formulation of King and Ross puts the  scale of the
family symmetry breaking at GUT scale or above, at which scale any
Higgs-like particles produced are too heavy and  decouple from low
energy dynamics or observability.  If we relax the need to tie the
$SU(3)_f$ breaking scale to GUT scale or above, we may find a
variable $SU(3)_f$ breaking scale that puts its physical scalar
particle spectrum within  experimental reach. \\

The new scalar particles to be observed, have peculiar properties.
They carry family labels (family charge), being triplets or
anti-triplets of $SU(3)_{f}$ like the quarks and leptons. The new
scalars are however singlets of the standard model $SU(2)_{L}
\otimes U(1)_{Y}$, meaning  the new scalar particles have zero
electric charge and zero weak hypercharge. Their masses are
completely unknown  and our interest in this paper, is to
ascertain the physical spectrum of these peculiar family symmetry
scalar particles, and what Higgs-like masses they may have. The
hope is that if these masses are not too high, the ongoing LHC
experiments at CERN may see these scalar bosons or set limits on
them. \\

Before we plunge into these concerns however, we  review in
section 2,  the salient features of the King-Ross formulation of
the fermion family symmetry model, built on GUTs and
supersymmetry, to be contrasted later with a non-Guts and
non-supersymmetric path we followed.  \\

\section{The King-Ross Family Symmetry Model of Fermion masses}
 A starting point is that King and Ross  formulated the fermion
mass problem into an effective field theory.  Such a formulation
requires that we regard the standard  model as a good theory
capable of  correctly describing electroweak physics but only
within a certain limited energy scale (the cut-off scale
$\Lambda_c$ of the standard model).  The physics of fermion mass
replication and flavor mixing is specifically believed  to lie
beyond  scale $\Lambda_c$. Rather at some high energy scale $M >
\Lambda_c$, it is believed that some more exact or fundamental
theory of electroweak physics exists, that provides a direct or
natural explanation for the fermion mass patterns and mixing
observed at standard model level. What this scale M is, and the
dynamics  there become  matters of conjecture and model building
\\

Whatever other details of the dynamics of scale M, it is believed
that an  fermion family symmetry such as $SU(3)_f$  exists there.
The symmetry is exact and fermion family triplets like $(u, c, t)$
or $(d,s,b)$ or $(e, \mu. \tau)$ are all massless there, and
correspond to triply degenerate  states $U^i, D^i, E^i$, with $ i
= 1,2,3$ running through each family. The family symmetry is
however subsequently broken, at or below  scale M. \\

King and Ross  chose scale M to be as high or higher than the GUT
scale of $10^{16}$ GeV which means that besides a gauged fermion
family symmetry $SU(3)_f$ at scale M, there are other gauge
symmetries there particularly GUT SO(10). The fields present at
such high energy scale are taken to be a variety of heavy
particles (bosons and fermions) along side the light fermions and
Higgs boson H of the standard model. \\

For the dynamics at scale M,  King and Ross adopt the Froggatt
-Nielsen model [7] where one has a variety of tree graphs mediated
by heavy messenger particles (bosons and fermions) coupled
directly to standard model fermions as external legs. When these
heavy messenger particles are later integrated out, one is led to
the low energy effective Lagrangian or effective Yukawa
couplings of the standard model, and thence  to fermion mass matrices.   \\

The model of King and Ross is chosen to be supersymmetric, and the
supersymmetry played two  central roles in the results obtained by
King and Ross. First the  Supersymmetry tied in a scalar potential
needed to break the family symmetry,  this potential being some
F-term of a  scalar superpotential given generally by [8]:

\begin{equation}\label{eq: ndili2}
W(\varphi_i) = \sum_i a_i \varphi_i  + \frac{1}{2} \sum_{ij}
m_{ij} \varphi_i \varphi_j  + \frac{1}{6} \sum_{ijk} y_{ijk}
\varphi_i \varphi_j \varphi_k +  \frac{1}{24} \sum_{ijkl}g_{ijkl}
\varphi_i \varphi_j \varphi_k \varphi_l +   ......\rightarrow
\infty
\end{equation}
where $\varphi_i$ is the scalar field component of the ith left
chiral superfield in the system, and $i, j, k, l ...$ run over all
left  chiral superfields in the system. Because one is dealing
with effective theory situation where  operators of higher
dimensions are permitted, the $W(\varphi_i)$ can run beyond $n \le
3 $ in powers of $\varphi_i$ normally required for renormalizable theories. \\

Every term of $W(\varphi_i)$ is required to be $SU(3)_f$
invariant. The first term   $W_1(\varphi_i) = a_i \varphi_i $ is
present only if the system contains singlet scalar fields of
$SU(3)_f$. The King-Ross model specifically   admitted a number of
such (heavy) singlet  scalar fields at scale M,  denoted
generically by X, and assigned  the role of generating F-terms and
vacuum alignment of other  scalar fields (triplets and
anti-triplets $\phi_3, \bar{\phi}_3, \phi_{23}, \bar{\phi}_{23} $
) used to break $SU(3)_f$ in a desired pattern.  Sample terms of
$W(\varphi_i)$ used to align the scalar fields  and achieve
spontaneous symmetry breaking in specific F-flat directions
include the following:
\begin{equation}\label{eq: ndili7}
W_3(\varphi_i) = X(\bar{\phi}_3^i \phi_{3i} - \mu^2)
\end{equation}
which generates the F-term $F_X = \bar{\phi}_3^i\phi_{3i} - \mu^2$
and the alignment of $\phi_3$ and $\bar{\phi}_3$ from the
minimization $|F_X| = \bar{\phi}_3^i\phi_{3i} - \mu^2 = 0$. Also
\begin{equation}\label{eq: ndili8}
W_{34}(\varphi_i) = X_1(\bar{\phi}_3^i \phi_{2i}) +
X_2(\bar{\phi}_{23}^i \phi_{3i}\bar{\phi}_{23}^j \phi_{2j} -\mu^4)
+ X_3 (\bar{\phi}_{23}^i \phi_{23i}) + ..
\end{equation}
whose separate terms align  $\phi_2$ relative to $\bar{\phi}_3$ ;
$\phi_{23}$ relative to $\phi_2$, and $\phi_{23}$ relative to
$\bar{\phi_{23}}$, etc. The end result is the following specific
alignment used by King-Ross to break  the fermion family symmetry
$SU(3)_f \rightarrow SU(2)_f \rightarrow 0$ :
\begin{equation}\label{eq: ndili1A}
 \phi_{3}  =   \left(
\begin{array}{c}
0 \\ 0 \\ a_3
\end{array}
\right) ;
 \bar{\phi_{3}}  =
(0,  0,  a_3) ;
 \phi_{23}  =   \left(
\begin{array}{c}
0 \\ b \\ b
\end{array}
\right) ;
 \bar{\phi_{23}}  =
(0, b. -b)
\end{equation}
It is with these particular  alignments, and the GUT- $SU(3)_f$
scale M dynamics  that King and Ross obtained their result that
fermion $SU(3)_f$ family symmetry  could be  the source of the
observed hierarchy of fermion masses and flavor mixing.  Notably
however, because the $SU(3)_f$ symmetry is broken at  so high
scale $M \geq $ GUT,  no lower energy scale observable signature
survives in the King-Ross model,  neither the very existence of
the family symmetry $SU(3)_f$, nor its spontaneous breaking. Only
the indirect inference exists through  the equally high scale
Froggatt-Nielsen heavy messenger particles, that the observed
fermion mass pattern may be due to the existence and spontaneous
breaking of a fermion family symmetry. It is against this
background that we consider a modified approach to the problem of
$SU(3)_f$ fermion family symmetry, and what signatures we may find for it.  \\

\section{A re-formulation of the fermion $SU(3)_f$ family symmetry model}
We retain the same effective theory frame work of King and Ross
with  a scale M that we treat as a variable,  not necessarily as
high as GUT scale.  We focus on $SU(3)_f$ as our main gauge
symmetry group at scale M.  We assume that this $SU(3)_f$ is
broken sequentially as before by the same scalar fields listed in
equation (4) and  aligned similarly,  but not necessarily arising
from any F-term or any assumed supersymmetry. In fact we seek to
construct directly [9] two separate scalar potentials $V(\phi_3,
\bar{\phi}_3)$ and $V(\phi_{23}, \bar{\phi}_{23})$ that we take to
be responsible for the two stage breaking : $SU(3)_f \rightarrow
SU(2)_f \rightarrow 0$.  They create  Higgs-like physical
particles we  can monitor or tag through varying scale M. With our
minimal assumptions, we can still construct an  effective Yukawa
Lagrangian weighted by our scale M, in  a manner similar to the
King-Ross model. Specifically, we write down the following
effective Yukawa couplings in our model
\begin{equation}\label{eq: ndili9a}
L_Y = \frac{1}{M^2} \bar{\phi}_3^i \psi_i \bar{\phi}_3^j \psi^c_j
H + \frac{1}{M^2}\bar{\phi}_{23}^i \psi_i \bar{\phi}_{23}^j
\psi^c_j H  + ...
\end{equation}
 These terms can generate fermion  masses exactly as before. \\

Leaving aside the issue of fermion masses, we focus on the scalar
potentials that break the $SU(3)_f$ symmetry, and the scalar mass
spectrum they may produce at variable scale M. \\

\section{The potential $V(\phi_3, \bar{\phi}_3)$ and $SU(3)_f \rightarrow SU(2)_f$ Breaking}
Up to fourth order we write $V(\phi_3, \bar{\phi}_3)$  as follows:
\begin{eqnarray}\label{eq: ndili7d}
V(\phi_{3} , \bar{\phi_{3}}) &=& \mu_1^2 \phi_3^\dag\phi_3  +
\mu_2^2\bar{\phi_3}^\dag\bar{\phi_3} +
\mu_{12}^2\phi_3^\dag\bar{\phi_3} +  \mu_{12}^{\star
2}\bar{\phi_3}^\dag\phi_3 \nonumber\\
 & & {} + \lambda_1(\phi_3^\dag\phi_3)^2 +
 \lambda_2(\bar{\phi_3}^\dag\bar{\phi_3})^2 +
 \lambda_3(\phi_3^\dag\phi_3)(\bar{\phi_3}^\dag\bar{\phi_3}) \nonumber\\
 & & {} + \lambda_4(\bar{\phi_3}^\dag\phi_3)(\phi_3^\dag
 \bar{\phi_3}) + \frac{\lambda_5}{2}(\phi_3^\dag\bar{\phi_3})^2 +
 \frac{\lambda_5^\star}{2}(\bar{\phi_3}^\dag\phi_3)^2 \nonumber\\
 & & {} + \frac{\lambda_6}{2}(\phi_3^\dag\phi_3)(\phi_3^\dag\bar{\phi_3}) +
 \frac{\lambda_6^\star}{2}(\bar{\phi_3}^\dag\phi_3)(\phi_3^\dag\phi_3)
 \nonumber\\
 & & {} +
 \frac{\lambda_7}{2}(\bar{\phi_3}^\dag\bar{\phi_3})(\phi_3^\dag\bar{\phi_3})
 +
 \frac{\lambda_7^\star}{2}(\bar{\phi_3}^\dag\phi_3)(\bar{\phi_3}^\dag\bar{\phi_3})
\end{eqnarray}

Requiring the potential to be hermitian means the unknown
parameters $\mu_1^2 , \mu_2^2, \lambda_1, \lambda_2, \lambda_3,
\lambda_4 $ are all real, while parameters $\mu_{12}^2 ,
\lambda_5, \lambda_6, \lambda_7$ are all complex. We can
additionally require $V(\phi , \bar\phi)$ to be symmetric with
respect to its two conjugate fields. This  leads to our putting
$\mu_1^2 = \mu_2^2 ; \lambda_1 = \lambda_2; $ and $\lambda_5 ,
\mu_{12}^2 $ real, while we drop $\lambda_6$ and $\lambda_7$ terms
as not tenable. We then rewrite  equation (6) obtaining :
\begin{eqnarray}\label{eq: ndili7e}
V(\phi_3, \bar{\phi_3}) &=&  \frac{\mu^2}{2}(\phi_3^\dag\phi_3 +
\bar{\phi_3}^\dag\bar{\phi_3}) +
\frac{\mu_{12}^2}{2}(\phi_3^\dag\bar{\phi_3}  +
\bar{\phi_3}^\dag\phi_3) \nonumber\\
& & {} + \frac{\lambda}{2} \left[(\phi_3^\dag\phi_3)^2  +
(\bar{\phi_3}^\dag\bar{\phi_3})^2\right] +
\lambda_3(\phi_3^\dag\phi_3)(\bar{\phi_3}^\dag\bar{\phi_3})
\nonumber\\
& & {} +
\lambda_4(\bar{\phi_3}^\dag\phi_3)(\phi_3^\dag\bar{\phi_3}) +
\frac{\lambda_5}{2}\left[(\phi_3^\dag\bar{\phi_3})^2 +
(\bar{\phi_3}^\dag\phi_3)^2\right]
\end{eqnarray}

where $\mu^2/2 = \mu_1^2 = \mu_2^2 $; and $\lambda/2 = \lambda_1 =
\lambda_2 $. \\

The scalar fields $\phi_3$ and $\bar{\phi}_3 $ are both complex
and can be parameterized in general as follows :  \\
\begin{equation}\label{eq: ndili3}
 \phi_{3}  =   \left(
\begin{array}{c}
\eta_1 + i\eta_2 \\ \eta_3 + i\eta_4 \\ \eta_5 + i\eta_6
\end{array}
\right) ; \langle \phi_3 \rangle = \left(
\begin{array}{c}
0 \\ 0 \\ a_3
\end{array}
\right)
\end{equation}

\begin{equation}\label{eq: ndili4}
\bar{\phi_{3}}  = ( \eta_1 - i\eta_2 ,  \eta_3 - i\eta_4 , \eta_5
- i\eta_6 )  ; \langle \bar{\phi_3} \rangle = ( 0 ,  0 , a_3)
\end{equation}
Similarly:
\begin{equation}\label{eq: ndili5}
 \phi_{23}  =   \left(
\begin{array}{c}
\eta_7 + i\eta_8 \\ \eta_9 + i\eta_{10} \\ \eta_{11} + i\eta_{12}
\end{array}
\right) ; \langle \phi_{23} \rangle = \left(
\begin{array}{c}
0 \\ b \\ b
\end{array}
\right)
 \end{equation}

\begin{equation}\label{eq: ndili6}
 \bar{\phi_{23}}  =
( \eta_7 - i\eta_8 , \eta_9 - i\eta_{10} ,\bar{\eta}_{11} -
i\eta_{12}) ; \langle \bar{\phi_{23}} \rangle =  ( 0 , b , -b )
 \end{equation}

We obtain from equations (8) and (9) the following final form for
$V(\phi_3, \bar{\phi}_3)$:
\begin{eqnarray}\label{eq: ndili7f}
V(\eta_1, \eta_2...\eta_6) &=&  \mu^2(\eta_1^2 + \eta_2^2 +
\eta_3^2 + \eta_4^2 + \eta_5^2 + \eta_6^2) + \mu_{12}^2(\eta_1^2 -
\eta_2^2 + \eta_3^2 - \eta_4^2 + \eta_5^2 - \eta_6^2) \nonumber\\
& & {} + 2\lambda(\eta_1^2 + \eta_2^2 + \eta_3^2 + \eta_4^2 +
\eta_5^2 + \eta_6^2)^2 \nonumber\\
& & {} + \lambda_4\left[(\eta_1^2 - \eta_2^2 + \eta_3^2 - \eta_4^2
+ \eta_5^2 - \eta_6^2)^2 + 4(\eta_1\eta_2 + \eta_3\eta_4 +
\eta_5\eta_6)^2\right] \nonumber\\
& & {} + \lambda_5\left[(\eta_1^2 - \eta_2^2 + \eta_3^2 - \eta_4^2
+ \eta_5^2 - \eta_6^2)^2 - 4(\eta_1\eta_2 + \eta_3\eta_4 +
\eta_5\eta_6)^2\right]
\end{eqnarray}
where we find $\lambda_3$ not a different coupling from $
\lambda$ and so have combined the two terms. \\

We are ready to analyze the potential (12) to find the physical
spectrum and masses of the scalar particles produced in this first
stage breaking of $SU(3)_f$. We use equation (4) or equations (8)
and (9), as what defines the vacuum of the potential. We take the
first and second derivatives of equation (12) evaluated at the
vacuum, and obtain
the following constraint equations on the $\mu_i , \lambda_i $ parameters. \\
\begin{equation}\label{eq: ndili4A}
\frac{\partial V}{\partial \eta_i}|_{vac} = 0 ; i = 1,2,3,4,6.
\end{equation}
\begin{equation}\label{eq: ndili4B}
\frac{\partial V}{\partial \eta_5}|_{vac} = \mu^2 + \mu_{12}^2 +
4\lambda a_3^2 + 2\lambda_4 a_3^2 + 2 \lambda_5 a_3^2 = 0
\end{equation}
\begin{equation}\label{eq: ndili4C}
\frac{1}{2}\frac{\partial^2 V}{\partial \eta_1^2}|_{vac} =
M_{11}^2 = \mu^2 + \mu_{12}^2 + 4\lambda a_3^2 + 2\lambda_4 a_3^2
+ 2 \lambda_5 a_3^2
\end{equation}
\begin{equation}\label{eq: ndili4D}
\frac{1}{2}\frac{\partial^2 V}{\partial \eta_3^2}|_{vac} =
M_{33}^2 = \mu^2 + \mu_{12}^2 + 4\lambda a_3^2 + 2\lambda_4 a_3^2
+ 2 \lambda_5 a_3^2
\end{equation}
\begin{equation}\label{eq: ndili4E}
\frac{1}{2}\frac{\partial^2 V}{\partial \eta_5^2}|_{vac} =
M_{55}^2 = \mu^2 + \mu_{12}^2 + 12\lambda a_3^2 + 6\lambda_4 a_3^2
+ 6 \lambda_5 a_3^2
\end{equation}
\begin{equation}\label{eq: ndili4F}
\frac{1}{2}\frac{\partial^2 V}{\partial \eta_2^2}|_{vac} =
M_{22}^2 = \mu^2 - \mu_{12}^2 + 4\lambda a_3^2 - 2\lambda_4 a_3^2
- 2 \lambda_5 a_3^2
\end{equation}
\begin{equation}\label{eq: ndili4G}
\frac{1}{2}\frac{\partial^2 V}{\partial \eta_4^2}|_{vac} =
M_{44}^2 = \mu^2 - \mu_{12}^2 + 4\lambda a_3^2 - 2\lambda_4 a_3^2
- 2 \lambda_5 a_3^2
\end{equation}
\begin{equation}\label{eq: ndili4H}
\frac{1}{2}\frac{\partial^2 V}{\partial \eta_6^2}|_{vac} =
M_{66}^2 = \mu^2 - \mu_{12}^2 + 4\lambda a_3^2 + 2\lambda_4 a_3^2
- 6 \lambda_5 a_3^2
\end{equation}
\begin{equation}\label{eq: ndili4h}
\frac{1}{2}\frac{\partial^2 V}{\partial \eta_i \partial
\eta_j}|_{vac} = M_{ij}^2 =  0 ; i \ne j
\end{equation}

We deduce from equations (14) - (16) that $M_{11}^2 = M_{33}^2 =
0, $ and that combining this with equation (17) we obtain that
\begin{equation}\label{eq: ndili4I}
M_{55}^2 = 4 a_3^2(2\lambda + \lambda_4 + \lambda_5 )  \geq 0
\end{equation}
Equations (15) - (17) are seen to define one group of three
particles (a family triplet),  while equations (18) - (20) define
another group of particles (a family anti-triplet). The symmetry
broken by $V(\phi_3 , \bar{\phi_{3}})$ is $SU(3)_f \rightarrow
SU(2)_f $ and we expect five massless Nambu Goldstone bosons two
of which are already identified by $M_{11}^2 = M_{33}^2 = 0$.
Similarly treating the second group of particles, we set $M_{22}^2
= M_{44}^2 = 0 $ which  leads to two more Nambu Goldstone bosons
and a residual mass:
\begin{equation}\label{eq: ndili4K}
M_{66}^2 = 4a_3^2( \lambda_4 - \lambda_5 )  \geq 0
\end{equation}
Looking at equation (12) we see that $\lambda_4$ and $\lambda_5$
essentially define the same coupling constant and we can set
$\lambda_4 = \lambda_5 $. Then $M_{66}^2 = 0 $ also, yielding the
required fifth Nambu Goldstone boson, and a final third generation
heavy neutral  physical particle $H_{5}$ of mass :
\begin{equation}\label{eq: ndili4L}
M_{55}^2 = 8 a_3^2(\lambda + \lambda_4 )
\end{equation}
related to our field $\eta_5$ by : $H_{5} = \eta_5 - a_3 $ since
$\langle H_5 \rangle = \langle \eta_5 \rangle - a_3 = 0$. We can
on the basis of equation (24), attempt to estimate the mass of
this Higgs-like particle $H_5$ produced in  this first $SU(3)_f$
symmetry breaking.  We argue as follows.  \\

First we argue that if M is the scale at which the $SU(3)_f
\rightarrow SU(2)_f$ breaking occurred, we can take the vev
parameter $a_3$  of the symmetry breaking as of the same order of
magnitude as scale M.  That is, we put $a_3 \approx M$ in equation
(24). \\

Next the parameters $\lambda $ and $\lambda_4$ in equation (24),
are both quartic scalar field couplings at the high energy scale M
where the potential $V(\phi_3, \bar{\phi}_3)$ operates. We can
assume that $\lambda \approx \lambda_4 = \lambda_M $ where
$\lambda_M$ is a running coupling constant at scale M for a
general $\lambda \phi^4$ system.  \\

Equation (24) becomes rewritten as:
\begin{equation}\label{eq: ndili4L}
M_{55}^2 = 8 M^2\lambda_M
\end{equation}

Next we  use the renormalization group  to relate the high scale
coupling $\lambda_M$ to a the standard model $\lambda_{SM}$ of
Higgs model at electroweak scale.  To one loop order the RG
relationship is [10]:
\begin{equation}\label{eq: ndili4p4}
\lambda_M = \frac{\lambda_{SM}}{1 - \frac{3 \lambda_{SM}}{16
\pi^2} \ln\frac{M}{\mu}}
\end{equation}
where  $\mu$ is the electroweak scale taken here to be  the $M_Z =
91.2$ GeV  $\approx 10^2 $ GeV. Taking $\lambda_{SM}$ to be very
small and perturbative based on indications from the standard
model, we find even for large values of scale M that we can write
the relationship equation (26) in the approximate form:
\begin{equation}\label{eq: ndili4p5}
\lambda_M = \lambda_{SM}[1 + a \lambda_{SM}]
\end{equation}
where $a = (3/(16 \pi^2))In(M/\mu)$ \\

Plugging equation (27) into (25) we obtain:
\begin{equation}\label{eq: ndili4L}
M_{55}^2 = 8 M^2 \lambda_{SM}[1 + a \lambda_{SM}]
\end{equation}
Next, we write down the standard model equation that relates
$\lambda_{SM}$ to standard model Higgs mass. It is:
\begin{equation}\label{eq: ndili4hM}
M_{Higgs}^2 =  2\lambda_{SM}v^2
\end{equation}
where v = 246 GeV. \\

 Eliminating $\lambda_{SM}$ between equations
(28) and (29) we finally obtain an equation relating the standard
model Higgs mass and  the new Higgs-like particle $M_{55}$
produced in the breaking of $SU(3)_f \rightarrow SU(2)_f$.  The
relation is:
\begin{equation}\label{eq: ndili4h}
\frac{M_{55}^2}{M^2} =  8 \frac{M_H^2}{v^2}  + 4 a
\left(\frac{M_H^2}{v^2} \right)^2
\end{equation}
This is our main finding, namely that if $SU(3)_f$ fermion family
symmetry exists at some high energy scale M where it is
spontaneously broken by some heavy scalar fields, the ratio of the
Higgs-like particle produced  to the symmetry breaking scale M, is
related in a distinct way to a similar ratio between the standard
model Higgs mass and standard electroweak breaking scale (vev) v.
Thus if a Higgs boson should turn up in the ongoing LHC
experiments  with a measured mass $M_H$, one can use equation (30
) to  estimate what $SU(3)_f$ breaking Higgs-like mass to expect
at a higher energy scale M.  \\

\section{The second stage $V(\phi_{23}, \bar{\phi}_{23})$
Breaking}
 We consider next the complementary  potential
$V(\phi_{23}, \bar{\phi}_{23})$ and the spontaneous $SU(2)_f
\rightarrow $ nothing  symmetry breaking it induces. The
potential is given by:  \\
\begin{eqnarray}\label{eq: ndili7j}
V(\phi_{23}, \bar{\phi_{23}}) &=& \frac{{\mu^\prime}^2}
{2}(\phi_{23}^\dag\phi_{23} + \bar{\phi_{23}}^\dag\bar{\phi_{23}})
+ \frac{{\mu_{12}^\prime}^2}{2}(\phi_{23}^\dag\bar{\phi_{23}}  +
\bar{\phi_{23}}^\dag\phi_{23}) \nonumber\\
& & {} + \frac{\lambda^\prime}{2}
\left[(\phi_{23}^\dag\phi_{23})^2 +
(\bar{\phi_{23}}^\dag\bar{\phi_{23}})^2\right] +
\lambda_3^\prime(\phi_{23}^\dag\phi_{23})(\bar{\phi_{23}}^\dag\bar{\phi_{23}})
\nonumber\\
& & {} +
\lambda_4^\prime(\bar{\phi_{23}}^\dag\phi_{23})(\phi_{23}^\dag\bar{\phi_{23}})
+
\frac{\lambda_5^\prime}{2}\left[(\phi_{23}^\dag\bar{\phi_{23}})^2
+ (\bar{\phi_{23}}^\dag\phi_{23})^2\right]
\end{eqnarray}

or using equations (10) and (11) we rewrite the potential  as:
\begin{eqnarray}\label{eq: ndili7k}
V(\eta_7, \eta_8, \eta_9, \eta_{10}, \eta_{11},  \bar{\eta}_{11},
\eta_{12}) &=& {\mu^\prime}^2 \left[\eta_7^2 + \eta_8^2 + \eta_9^2
+ \eta_{10}^2 + \frac{1}{2}(\eta_{11}^2 + \bar{\eta}_{11}^2) + \eta_{12}^2 \right]  \nonumber\\
& & {} + {\mu_{12}^\prime}^2(\eta_7^2 -
\eta_8^2 + \eta_9^2 - \eta_{10}^2 + \eta_{11}\bar{\eta}_{11} - \eta_{12}^2) \nonumber\\
& & {} + \frac{\lambda^\prime}{2}(\eta_7^2 + \eta_8^2 + \eta_9^2 +
\eta_{10}^2 + \eta_{11}^2 + \eta_{12}^2)^2 \nonumber\\
& & {} +  \frac{\lambda^\prime}{2}(\eta_7^2 + \eta_8^2 + \eta_9^2
+ \eta_{10}^2 + \bar{\eta}_{11}^2 + \eta_{12}^2)^2 \nonumber\\
& & {} + \lambda_3^\prime (\eta_7^2 + \eta_8^2 + \eta_9^2 +
\eta_{10}^2 + \eta_{11}^2 + \eta_{12}^2) \times
\nonumber\\
& & {} (\eta_7^2 + \eta_8^2 + \eta_9^2 + \eta_{10}^2 +
\bar{\eta}_{11}^2 + \eta_{12}^2) \nonumber\\
 & & {} + \lambda_4^\prime(\eta_7^2 - \eta_8^2 +
\eta_9^2 - \eta_{10}^2 + \eta_{11}\bar{\eta}_{11} - \eta_{12}^2)^2  \nonumber\\
& & {} + \lambda_4^\prime (2\eta_7\eta_8 + 2\eta_9\eta_{10}
+ \eta_{11}\eta_{12} + \bar{\eta}_{11}\eta_{12})^2 \nonumber\\
& & {} + \lambda_5^\prime ( \eta_7^2 - \eta_8^2 + \eta_9^2 -
\eta_{10}^2 + \eta_{11}\bar{\eta}_{11} - \eta_{12}^2)^2
\nonumber\\
& & {} - \lambda_5^\prime(2\eta_7\eta_8 + 2\eta_9\eta_{10} +
\eta_{11}\eta_{12} + \bar{\eta}_{11}\eta_{12})^2
\end{eqnarray}
It is this potential we analyze for the additional set of scalar
particles created by the $V(\phi_{23} , \bar{\phi}_{23})$.
 We find as follows:
\begin{equation}\label{eq: ndili4AA}
\frac{\partial V}{\partial \eta_i}|_{vac} = 0 ; i = 7, 8, 10, 12..
\end{equation}
\begin{equation}\label{eq: ndili4BB}
\frac{\partial V}{\partial \eta_9}|_{vac} = {\mu^\prime}^2 +
{\mu_{12}^\prime}^2 + 4b^2(\lambda^\prime + \lambda_3^\prime) = 0
\end{equation}
\begin{equation}\label{eq: ndili4BC}
\frac{\partial V}{\partial \eta_{11}}|_{vac} = {\mu^\prime}^2 -
{\mu_{12}^\prime}^2 + 4b^2(\lambda^\prime + \lambda_3^\prime) = 0
\end{equation}
\begin{equation}\label{eq: ndili4CC}
\frac{1}{2}\frac{\partial^2 V}{\partial \eta_7^2}|_{vac} =
M_{77}^2 = {\mu^\prime}^2 + {\mu_{12}^\prime}^2 +
4b^2(\lambda^\prime + \lambda_3^\prime)
\end{equation}
\begin{equation}\label{eq: ndili4DD}
\frac{1}{2}\frac{\partial^2 V}{\partial \eta_8^2}|_{vac} =
M_{88}^2 =  {\mu^\prime}^2 - {\mu_{12}^\prime}^2 +
4b^2(\lambda^\prime + \lambda_3^\prime)
\end{equation}
\begin{equation}\label{eq: ndili4EE}
\frac{1}{2}\frac{\partial^2 V}{\partial \eta_9^2}|_{vac} =
M_{99}^2 =  {\mu^\prime}^2 + {\mu_{12}^\prime}^2 +
8b^2(\lambda^\prime + \lambda_3^\prime) +4b^2(\lambda_4^\prime +
\lambda_5^\prime)
\end{equation}
\begin{equation}\label{eq: ndili4FF}
\frac{1}{2}\frac{\partial^2 V}{\partial \eta_{10}^2}|_{vac} =
M_{10,10}^2 = {\mu^\prime}^2 - {\mu_{12}^\prime}^2 +
4b^2(\lambda^\prime + \lambda_3^\prime + \lambda_4^\prime -
\lambda_5^\prime)
\end{equation}
\begin{equation}\label{eq: ndili4GG}
\frac{1}{2}\frac{\partial^2 V}{\partial \eta_{11}^2}|_{vac} =
M_{11,11}^2 = \frac{1}{2}{\mu^\prime}^2 + b^2(4\lambda^\prime +
2\lambda_3^\prime + \lambda_4^\prime + \lambda_5^\prime)
\end{equation}
\begin{equation}\label{eq: ndili4HH}
\frac{1}{2}\frac{\partial^2 V}{\partial \eta_{12}^2}|_{vac} =
M_{12,12}^2 = {\mu^\prime}^2 - {\mu_{12}^\prime}^2 +
4b^2(\lambda^\prime + \lambda_3^\prime)
\end{equation}
The only mixing found in the system is between fields $\eta_9$ and
$\eta_{11}$ where
\begin{equation}\label{eq: ndili4Hh}
\frac{1}{2}\frac{\partial^2 V}{\partial \eta_9 \partial
\eta_{11}}|_{vac} = M_{9,11}^2 =  M_{11,9}^2 = 2b^2(\lambda^\prime
+ \lambda_3^\prime - \lambda_4^\prime - \lambda_5^\prime)
\end{equation}

Before  discussing this equation, we draw some conclusions from
equations (34) - (41). Based on these equations, the fields are
seen again to divide into two groups: Fields $\eta_7, \eta_9,
\eta_{11}$ form a triplet while fields $\eta_8, \eta_{10},
\eta_{12}$ represent anti-triplet effect. Based on equation (34)
we deduce $M_{77}^2 = 0;$ while equation (38) becomes:
\begin{equation}\label{eq: ndili4e}
M_{99}^2 = 4b^2(\lambda^\prime + \lambda_3^\prime +
\lambda_4^\prime + \lambda_5^\prime) \geq 0
\end{equation}
Also based on equation (35) we see $M_{88}^2 = M_{12,12}^2 = 0 $
while equation (39) simplifies to :
\begin{equation}\label{eq: ndili3e}
M_{10,10}^2 = 4b^2(\lambda_4^\prime - \lambda_5^\prime )
\end{equation}
We have thus three massless particles $\eta_7, \eta_8, \eta_{12}$
we expect as Nambu Goldstone bosons for $SU(2) \rightarrow $
nothing. In addition if we make the assumption that
$\lambda_4^\prime \approx \lambda_5^\prime$ ( both being part of a
general scalar running quartic coupling stated below), we get a
fourth massless particle  $ \eta_{10}.$ \\

There remain two fields to account for: $\eta_9$ and $\eta_{11}$.
They appear to mix as indicated  by equation (42). If  $H_9$ and
$H_{11}$ are  two physical particles associated with $\eta_9$ and
$\eta_{11}$, then upon  diagonalization we find these particle
masses to be
\begin{equation}\label{eq: ndili3b}
M_{H_9}^2 = -\frac{1}{2}B + \frac{1}{2} \sqrt{B^2 - 4C}
\end{equation}
\begin{equation}\label{eq: ndili3j}
M_{H_{11}}^2 = -\frac{1}{2}B - \frac{1}{2} \sqrt{B^2 - 4C}
\end{equation}
where :
\begin{eqnarray}\label{eq: ndili7nq}
B & = & - (4b^2X_1 + \frac{1}{2} {\mu^\prime}^2 + b^2X_3)
\nonumber\\
 C & = & (4b^4X_1X_3 + 2b^2X_1{\mu^\prime}^2
 -4b^4X_2^2)\nonumber\\
 X_1 & = & \lambda_1^\prime + \lambda_3^\prime + \lambda_4^\prime
 + \lambda_5^\prime \nonumber\\
X_2 & = & \lambda_1^\prime + \lambda_3^\prime - \lambda_4^\prime
 - \lambda_5^\prime \nonumber\\
 X_3 & = & 4\lambda_1^\prime + 2\lambda_3^\prime + \lambda_4^\prime
 + \lambda_5^\prime
\end{eqnarray}

We can now try to  estimate the scale of these masses by arguing
variously as before.  First we argue  that all the couplings
$\lambda_1^\prime , \lambda_3^\prime , \lambda_4^\prime ,
\lambda_5^\prime $  are quartic scalar field couplings at the same
high energy scale $M^\prime$ where $SU(2)_f \rightarrow$ nothing
occurs. As such we can represent each of them by the same running
coupling $\lambda_{M^\prime}$ at scale $M^\prime$ of a $\lambda
\phi^4$ theory. Then we write:
\begin{eqnarray}\label{eq: ndili17nq}
 X_1 & = & 4 \lambda_{M^\prime} \nonumber\\
X_2 & = &  0 \nonumber\\
 X_3 & = & 8\lambda_{M^\prime}
\end{eqnarray}
Notably too, the interference term $M_{11,9}^2$ in equation (42)
drops out, leaving us to deal directly with equations  (40) and
(43). From equation (43)  we get
\begin{equation}\label{eq: ndili4A}
M_{99}^2 = 16b^2\lambda_{M^\prime}
\end{equation}
while from equation (40) we get:
\begin{equation}\label{eq: ndili4fG}
M_{11,11}^2 = \frac{1}{2}{\mu^\prime}^2 + 8 b^2\lambda_{M^\prime}
\end{equation}
Then similar to equation (27) from the renormalization group, we
can take:
\begin{equation}\label{eq: ndili4p7}
\lambda_{M^\prime} = \lambda_{SM}[1 + a^\prime \lambda_{SM}]
\end{equation}
where $a^\prime = (3/(16 \pi^2))In(M^\prime/\mu)$ This transforms
equation (49) and (50) to :
\begin{equation}\label{eq: ndili4C}
M_{99}^2 = 16b^2\lambda_{SM}[1 + a^\prime \lambda_{SM}]
\end{equation}
\begin{equation}\label{eq: ndiliF}
M_{11,11}^2 = \frac{1}{2}{\mu^\prime}^2 + 8 b^2\lambda_{SM} [ 1 +
a^\prime \lambda_{SM}]
\end{equation}

Finally we bring in standard model Higgs mass as a reference scale
related to $\lambda_{SM}$ and v = 246 GeV by equation (29),
 $M_{Higgs}^2 = 2v^2\lambda_{SM}.$  Also we can take the vacuum expectation
value b of the  $SU(2)_f \rightarrow $  nothing breaking at scale
$M^\prime $ as comparable in value to $M^\prime. $ That is we put
$ b = M^\prime$. Then equations (52) and (53) take the final form
\begin{equation}\label{eq: ndili4F}
\frac{M_{99}^2}{{M^{\prime}}^2} =  8 \frac{M_H^2}{v^2}  + 4
a^\prime \left(\frac{M_H^2}{v^2} \right)^2
\end{equation}
\begin{equation}\label{eq: ndiliF1}
2M_{11,11}^2  - M_{99}^2 = c
\end{equation}
where  c is a constant. We see equation (54) is the same formula
as equation (30), but holds at scale  $M^\prime $ where the
secondary spontaneous breaking $SU(2)_f \rightarrow $ nothing,
takes place. \\

\section{Summary and Conclusions}
We summarize our results and draw conclusions. We set out to
examine the physical spectrum and masses of new scalar particles
that should exist if continuous family symmetries such as
$SU(3)_f$ symmetry proposed by King et. al.[1-6]  exist and become
completely broken down spontaneously above the electroweak scale.
Our analysis shows that besides the expected numbers of massless
Numbu Goldstone bosons, some  massive scalar bosons  $H_5 , H_9,
H_{11}$, much heavier than the Higgs boson should also exist. We
obtained the main result stated in equations (30) and (54),  that
the ratio of the mass of these  heavy Higgs bosons  to the scale M
at which their $SU(3)_f$ symmetry is broken,  can be determined by
the standard model Higgs boson mass and the electroweak scale v.
We consider that this formula can be a guide in searching for
heavy Higgs bosons as evidence of $SU(3)_f $ type fermion family symmetry. \\

These particles are postulated to carry no electroweak  quantum
numbers and no color charges. They carry only family  or
generation quantum number through which they couple to fermions
that also carry generation numbers. If the coupling of these heavy
scalar bosons is by the gauge principle as we assumed, then new
gauge bosons other than the electroweak and gluonic gauge bosons
must exist to mediate the family (generation) force at high
energies.  \\

By way of other competing new Higgs bosons, we mention other
models that propose new scalar particles in the electroweak
system. We have in particular, the two Higgs doublet  model 2HDM
[11,12], and the minimal supersymmetric model MSSM [11,12], the
latter being however only a special case of 2HDM. Each of these
two models independently predicts a total of five new Higgs-like
scalar particles compared to the one Higgs particle of the
standard model. The five particles are usually denoted by: $H^\pm,
A^o, H^o, h^o $ of which $h^o$ is believed to be the lightest.
Their masses are however unknown except for various   bounds and
limits placed from experimental searches or unseen  decay rates
[13- 16]. Many workers in these models think  however, that none
of the five particles is likely to have mass exceeding 700 GeV.
This is likely to be  much lighter than our own particles $H_5 ,
H_9, H_{11}$, unless the scale M at which our $SU(3)_f$ family
symmetry is actually broken turns out not to be exceedingly high. \\ \\

\textbf{References :} \\

1. S. F. King and G. G. Ross, Phys. Letters B520 (2001) 243. \\

2.  S. F. King and G. G. Ross, Phys. Letters B574 (2003) 239. \\

3. Ivo de Medeiros Varzielas, University of Oxford Ph.D.
   thesis, \\ hep-ph/0801.2775 (2008)  \\

4. Ivo de Medeiros Varzielas and G. G. Ross, Nucl. Phys. B733
(2006) 31 \\

5. G. G. Ross and L. Velasco-Sevilla, Nucl. Phys. B653 (2003) 3 \\

6. S. F. King and M. Malinsky, JHEP 0611 : 071, (2006)  \\

7. C. D. Froggatt and H. B. Nielsen, Nucl. Phys. B147 (1979) 277
\\

8. A. Signer, ABC of SUSY, hep-ph/0905.4630 (2009) \\

9. L. F.  Li, Phys. Rev. D9 (1974) 1723 \\

10. T. P. Cheng and L. F. Li, Gauge theory of elementary particle
    physics, Clarendon Press Oxford, 1984. \\

11. J. L. Diaz-Cruz and A. Mendez, Nucl. Phys.  B380 (1992) 39 \\

12. J. F. Gunion and H. Haber, Nucl. Phys.  B272 (1986) 1 \\

13. S. Heinemeyer, Higgs Physics at the LHC: Some theory aspects,
arXiv:hep-ph/08072514 (2008). \\

14. Shinya Kanemura et. al. Distinctive Higgs signals of a type II
2HDM at the LHC : arXiv:hep-ph/09010204 (2009). \\

15. J. Ellis, Outlook for charged Higgs Physics,
arXiv:hep-ph/09011120
(2009) \\

16. B. Dudley and C. Kolda, Constraining the charged Higgs mass in
the MSSM: arXiv:hep-ph/09013337 (2009).

\end{document}